\documentclass[12pt,preprint]{aastex}

\shorttitle{Gamma-Ray Burst Hubble Diagram}
\shortauthors{Schaefer}

\received{2002 July 30}
\begin{document}

\title{Gamma-Ray Burst Hubble Diagram to $z=4.5$}

\author{Bradley E. Schaefer}
\affil{Department of Astronomy, University of Texas,
    Austin, TX 78712}

\begin{abstract}

Gamma-ray bursts (GRBs) are tremendous explosions visible across most of
the Universe, certainly out to redshifts of $z=4.5$ and likely out to $z
\sim 10$.  Recently, GRBs have been found to have a roughly constant
explosive energy as well as to have two luminosity indicators (the
spectral lag time and the variability) that can be used to derive the
burst's luminosity distance from the gamma-ray light curve alone.  There
currently exists enough information to calibrate luminosity distances and
independent redshifts for nine bursts.  From these, a GRB Hubble diagram
can be constructed, where the observed shape of the curve provides a
record of the expansion history of our Universe.  The current nine burst
diagram is sparse, yet formal limits can be placed on the mass density of
a flat Universe.  This first GRB Hubble diagram provides a proof of
concept for a new technique in cosmology at very high redshifts.  With
the launch of the SWIFT satellite in 2003, we should get $\sim 120$
bursts to produce a Hubble diagram impervious to all effects of dust
extinction and out to redshifts impossible to reach by any other method.

\end{abstract}

\keywords{cosmology: cosmological parameters---cosmology: distance
scale---cosmology: observations---gamma rays: bursts}


\section{Introduction}

	As always in astronomy, the determination of distances is a
crucial and difficult problem.  Until recently, the distance scale of
GRBs was unknown by over 12 orders-of-magnitude.  In 1997, the discovery
of optical and radio counterparts \citep{cos97,van97,fra97} proved that
at least the long-duration bursters were at cosmological distances with
redshifts of $z \sim 1$.  The measurement of GRB redshifts requires deep
optical spectra, and to date only 24 redshifts are known for bursts with
unknown selection effects.

	If GRB distance indicators can be found that use only gamma-ray
data, then we can measure the demographics and cosmology of large and
well-understood samples of bursts.  Two such luminosity (and hence
distance) indicators have recently been proposed.  The first
\citep{nmb00} relates the burst luminosity ($L$) with the spectral lag
($\tau_{lag}$), which can be idealized as the time between peaks as
recorded at high and low photon energies.  The second \citep{frr00}
relates the luminosity with the variability ($V$), which is a specific
measure of the 'spikiness' of the burst light curve.  High luminosity
bursts have short lags and spiky light curves, while low luminosity
events have long lags and smooth light curves.  These two relations make
GRBs into 'standard candles', in the same sense as for Cepheids and
supernovae where an observed light curve property can yield the
luminosity and then the distance.

	The two luminosity indicators were originally proposed and
calibrated with 6 or 7 bursts.  The addition of 3 or 2 further bursts
(for a total of nine) have fallen on the original relations,
hence adding confidence in their utility.  Further, if both the $L /
\tau_{lag} $ and $L/V$ relations are true, then there must be a
particular $ \tau_{lag} / V$ relation, and this prediction has been
strongly confirmed with an independent sample of 112 BATSE bursts
\citep{sdb01}.  Finally, the very long lag bursts have been shown to be
very low luminosity as demonstrated by their concentration to the local
supergalactic plane \citep{nor02}.

	These luminosity indicators have been used to identify specific
bursts \citep{frr00} that are at redshifts of $z \sim 10$ as well as to
show that the star formation rate of the Universe is rising steadily
\citep{frr00,sdb01,lfr01} from $z \sim 2$ to $z>6$.  This {\it Letter}
reports on the construction of a Hubble diagram (a plot of luminosity
distance, $D_L$, versus redshift) as a means of measuring the expansion
history of our Universe.

\section{GRB Hubble Diagram}

	Only nine GRBs have the required information of red shift ($z$),
peak flux ($P$), lag time ($\tau_{lag}$), and variability ($V$).  These
data are collected in Table 1, along with the characteristic photon
energy ($E_{peak}$) and the observed luminosity ($L_{obs}$).  These nine
bursts were all detected by BATSE with redshifts measured
from optical spectra of either the afterglow or the host galaxy.  The
highly unusual GRB980425 (associated with supernova SN1998bw) is not
included because it is likely to be qualitatively different from the
classical GRBs.  Bursts with red shifts that were not recorded by BATSE
cannot (yet) have their observed parameters converted to energies and
fluxes that are comparable with BATSE data.  

	Simplistically, plots of $L_{obs}$ versus $\tau_{lag}$ and
$L_{obs}$ versus $V$ can calibrate the luminosity indicators, which then
can yield luminosity distances to each burst for plotting on a Hubble
Diagram and fitting to cosmological models.  In practice, there must be a
simultaneous chi-square minimization for the luminosity calibrations and
the cosmology, so as to avoid the effects of possible correlations
between the indicators and distance.

If the Earth is not along the central axis of the GRB's jet, then various
off-axis effects will change the observed properties. For light viewed
off-axis by angle $\theta$ from a jet moving at $\beta$ times the speed
of light, the transverse Doppler shift, redshift, and beaming will change
the various observed properties by powers of $B=(1- \beta )/(1- \beta cos
\theta )$.  The observed value of $E_{peak}$ will scale as $B$,
$\tau_{lag}$ will scale as $B^{-1}$, $V$ will scale as the inverse of a
time and hence as $B$, while $L$ will have a model dependent variation
that scales roughly as $B^3$.  For structured jets \citep{rlr02,zhm02},
the luminosity and bulk Lorentz factor can be parameterized as being
proportional to some positive power of $B$.  When viewed on-axis, the
bursts will presumably display fairly tight lag/luminosity and
variability/luminosity relations.  These relations can be converted to
similar relations involving observed off-axis quantities with an extra
factor of $B$ to some power. Fortunately, the observed distribution of
$E_{peak}$ is remarkably narrow for bursts of a given brightness
\citep{mal95,bra98}, and the removal of kinematic effects shows that the
on-axis $E_{peak}$ is virtually constant \citep{sch02}.  Thus, $E_{peak}$
should be proportional only to some power of $B$, and this relation can
be used to convert the $B$ dependency of the luminosity relations into a
$E_{peak}$ dependency.  Essentially, we can use the observed $E_{peak}$
value to get information about the off-axis angle and correct the
observed $\tau_{lag}$ and $V$ values to those which would be seen
on-axis.  In practice, the correction factor can be taken as some power
of $E_{peak}(1+z)/400keV$.  (The division by 400 keV is to minimize
correlations between the normalization constant and the exponent during
the fits.)  The exponents of the correction factors will be free
parameters which depend on the scenario and jet structure.

	With the off-axis correction, the luminosities can be derived by
fitting the relations $L \propto \tau_{lag,corr}^{\alpha_{lag}}$ and $L
\propto V_{corr}^{\alpha_{V}}$, where $\tau_{lag,corr} = \tau_{lag}
\times [E_{peak} (1+z)/400] ^{e_{lag}}$ and $V_{corr} = V \times
[E_{peak} (1+z)/400] ^{e_V}$.  Plots of $L_{obs}$ versus
$\tau_{lag,corr}$ and $L_{obs}$ versus $V_{corr}$ for the nine bursts
with known redshifts (see Figure 1) have slopes of $\alpha_{lag}$ and
$\alpha_V$, while the scatter in these plots will be minimized for the
best values of $e_{lag}$ and $e_V$.  This gives $\alpha_{lag} = -1.27 \pm
0.20$, $\alpha_V = 1.57 \pm 0.17$, $e_{lag} = 0.6 \pm 0.7$, and $e_V =
0.85 \pm 0.40$.  Thus:
\begin{equation}
L = 10^{50.03} [(\tau_{lag} \times \{E_{peak} (1+z)/400\} ^{0.6}]^{-1.27} ,
\end{equation}
\begin{equation}
L = 10^{55.32} [(V \times \{E_{peak} (1+z)/400\} ^{0.85}]^{1.57} ,
\end{equation}
where $L$ is in $erg ~s^{-1}$, $\tau_{lag}$ is in seconds, $V$ is
dimensionless, and $E_{peak}$ is in keV.  As independent checks, a plot
of $\tau_{lag,corr}$ versus $V_{corr}$ for 93 BATSE bursts will have a
slope of $\alpha_{lag} / \alpha_V$, and the scatter will be minimized for
the best values of $e_{lag}$ and $e_V$.  This gives $\alpha_{lag} /
\alpha_V = -0.7 \pm 0.2$, $e_{lag} = 0.0 \pm 0.3$, and $e_V = 1.1 \pm
0.3$; values which are consistent with the calibration from Figure 1.

	The lag and the variability produce two independent luminosity
values through Eqs 1 and 2.  These can be combined as weighted averages
(with the weights equal to the inverse square of the one-sigma errors) to
produce a combined luminosity ($L_{comb}$) and then $D_L$, as given in
Table 1.  A plot of $D_L$ versus $z$ is a Hubble diagram (Figure 2).

	Equations 1 and 2 are optimal for studies of GRBs where the
cosmology is not in question.  When the goal is to use the bursts as
cosmological markers, care must be taken to avoid circular logic.  In
particular, the previous paragraph assumed a specific cosmology
($\Omega_m =0.3$ for a flat Universe and $H_0 = 
65~{\rm km~s}^{-1}~{\rm Mpc}^{-1}$)  
for calculating the $D_L$ values as a function of $z$.  To avoid this
circularity, the fits must be performed with the cosmological parameters
as additional fit parameters.  That is, the model being fit to the data
will have cosmological plus GRB parameters that are to be simultaneously
determined by the minimization of the chi-square.  This will tie the
cosmology conclusions into GRB calibrations, but both will be well
constrained when many bursts are available. A correct procedure to find
the best value of $\Omega_m$ in a flat Universe is (1) to fix $\Omega_m$,
(2) derive $D_L$ and $L_{obs}$ for each burst for that cosmology, (3) fit
$L_{obs}$ as power laws of $\tau_{lag,corr}$ and $V_{corr}$, (4) use the
best fits to derive $L_{comb}$ and then $D_L$ for each burst, (5)
calculate chi-square by comparing the model and observed $D_L$ values,
(6) repeat steps 1-5 to identify the best fit $\Omega_m$ value as that
where chi-square is minimized, and (7) identify the one-sigma range where
chi-square is within unity of its minimum value.  By this procedure, the
lowest chi-square corresponds to $\Omega_m =0.05$ with the one-sigma
constraint $\Omega_m < 0.35$.  The three-sigma limit is larger than
unity, so this particular measure is not helpful for cosmology.  This new
result is completely independent of, yet in agreement with, those from
supernovae and other methods \citep{dun01}.  However, the current value
for $\Omega_m$ is not yet of high reliability because the number of
degree of freedom in the fit are small and the slope of the Hubble
diagram is dominated by just two high redshift events.  While this new
result does not seriously constrain current cosmology, it does act as a
demonstration of principle and a sign for the future.

\section{Comparison with Supernovae}

	A comparison with the supernova-based Hubble diagram is
inevitable.  GRBs have both advantages and disadvantages when compared to
supernovae.

	At least currently, supernovae have an advantage over GRBs for
the accuracy of the distance measurements for a single event.  
Well-observed nearby Type Ia supernovae have an RMS scatter about their
Hubble Diagram of 0.18 mag \citep{phi99} which translates into an
uncertainty in the log of the distance of 0.04.  GRBs currently have an
RMS scatter about their calibration curves of 0.35 and 0.20 in the logs
of lag and variability; which translates into uncertainties in the log of
the distance of 0.22 and 0.16 respectively.  These two independent
measures can be combined to give an uncertainty of 0.13 in the log of the
distance.  Thus, individual supernovae are roughly three times more
accurate as standard candles than are individual GRBs.

	The physics of gamma-ray emission from relativistic shocks in
GRBs is largely known (but not how to create the magnetic fields)
although the explosion scenario is still uncertain.  The basic physics of
supernovae is well known (although the Type Ia progenitors are still not
identified).  Supernovae can have their decline-rate versus luminosity
calibration from low-redshift events, and this allows the elegance of
being able to make the calibration substantially separate from the
cosmology.  Thus, supernovae are much better understood than are GRBs,
and this will be a substantial comfort for using them as standard
candles.  Nevertheless, it is unclear where this advantage for supernovae
will pay off.  Currently, the analyses of both supernova and GRB Hubble
Diagrams are entirely empirical with no contribution from theory, so a
better theoretical understanding of supernovae has not helped in any
specific manner.  Supernova will have better known evolution effects; yet
\citet{bra01} show that evolution is no problem for supernovae and their
argument applies identically to GRBs.

	Supernovae have substantial problems with a higher dust density
at high red shift \citep{tok99} and the possibility of grey dust
\citep{agu99}.  Gamma radiation suffers from no extinction.

	After a tremendous observational effort, the current supernova
Hubble diagram \citep{per99} extends only out to $z=0.97$. (SN1997ff at
$z \sim 1.7$ has very large uncertainties \citep{rie01} even without the
large corrections for gravitational lensing \citep{lei01,mgg01} and the
uncertainty of the supernova type.)  The dedicated SNAP satellite
\citep{sna02}, proposed for launch in 2008, will go out to $z=1.7$ with
exquisite accuracy in the light curves.  In contrast, the GRB Hubble
Diagram is already out to $z=4.5$ and likely it will be extended to $z
\sim 10$ \citep{frr00,lfr01} or farther \citep{lar00,brl02}.  With the
launch of the SWIFT satellite in 2003, the GRB Hubble Diagram will be
available for the redshift ranges $1 < z < 1.7$ and $1.7 < z \lesssim 10$
roughly five years before SNAP can extend the supernova Hubble Diagram
only to $1 < z < 1.7$.

	In summary, GRBs are $\sim 3 \times$ worse in accuracy than
supernovae, but this is traded-off for the lack of problems from
extinction, extension to $z \sim 10$, and an answer many years before
SNAP.  I do not think that the GRB Hubble Diagram will replace the
supernova Hubble Diagram.  The reason is that {\it both} diagrams will
have known and unknown systematic problems which will make results from
any {\it one} method not conclusive.  What is needed is the concurrence
of multiple independent methods.  So {\it both} GRB and supernova Hubble
Diagrams are needed for a confident result.

\section{Implications}

	Conceptually, the biggest question related to a construction of
the GRB Hubble diagram is the possibility of luminosity evolution.  
Fortunately, there are two strong reasons that any such effects must be
small.  First, theoretical explanations for the luminosity indicators
\citep{sch01,ion01,pla01} tie the relations to simple physics (involving
relativistic effects operating on any light source, the delay in light
travelling different paths, and the cooling of any body by radiation)
that are unrelated to properties (such as metallicity) that might evolve.  
Second, just as for the purely empirical supernova Hubble diagram
\citep{bra01}, any drift in population average properties \citep{lfr01}
is irrelevant as the calibrations will work for any individual event
whether near or far.  That is, it does not matter whether population
average properties (like luminosity) evolve because the distances are
derived for individual events that are each correctly calibrated.

	Within two years, the GRB situation will be greatly improved with
the dedicated SWIFT satellite \citep{geh00} currently scheduled for
launch in September 2003.  SWIFT is expected to get $\sim 100$ bursts per
year, of which $\sim 40 \%$ will have directly measured redshifts.  The
redshift will be difficult to measure for $z > 6$ bursts, yet it is
possible by locating the Lyman break with broad-band photometry or
spectroscopy extending into the near infrared.  In its nominal three-year
lifetime, we can get $\sim 120$ bursts with measured values for $z$, $P$,
$\tau_{lag}$, $V$, and $E_{peak}$.  This many bursts will allow for high
accuracy calibration of the luminosity indicators.  The large number of
bursts will also allow for the reduction of statistical errors (dominated
by the intrinsic scatter of the bursts) by a factor of $\sim 10$.  If the
bursts are divided into six redshift bins, each bin will have an error of
0.03 in the log of the distance.  This accuracy will, for example, yield
an uncertainty of $\sim 0.03$ in the derived value of $\Omega_m$ for a
flat Universe, even with no improvements in the luminosity indicators.  
So with SWIFT, the prospect is that we can produce an accurate GRB Hubble
diagram with $\sim 120$ bursts from $0.1<z<10$ by 2006.

	What can we learn from the GRB Hubble Diagram?  We can test the
predicted shift of the Universe from matter to dark-energy domination in
the range $1<z<2$.  We will also look for various predicted quintessence
\citep{wea01} and non-standard \citep{man02} effects in the range
$2<z<10$.  The typical sizes of these effects are 0.4 in the log of the
distance around a redshift of $z=3$.  The current paradigm of cosmology
(the new inflation perhaps with quintessence) is so new and untested that
surprises could easily await in the unknown regime of $z>1$.

\clearpage

\begin{deluxetable}{ccccccccc}
\tabletypesize{\scriptsize}
\tablecaption{GRB Luminosity Distances From Lags and Variabilities. \label{tbl-1}}
\tablewidth{0pt}
\tablehead{
\colhead{GRB} & \colhead{$z$}   & \colhead{$P$\tablenotemark{a}}   &
\colhead{$\tau_{lag}$\tablenotemark{b}} &
\colhead{$V$\tablenotemark{c}}  & \colhead{$E_{peak}$\tablenotemark{d}} & 
\colhead{$Log(L_{obs})$\tablenotemark{e}} &
\colhead{$Log(L_{comb})$\tablenotemark{f}} &
\colhead{$D_L$\tablenotemark{g}}
}
\startdata
970508	&	0.84	&	$1.2	\pm	0.1$	&	$0.307	\pm	0.065$	&	$0.0010	\pm	0.0010$	&	$137	\pm	14$	&	$50.89	\pm	0.04$	&	$50.70	\pm	0.39$	&	$4600	\pm	2100$	\\
970828	&	0.96	&	$4.9	\pm	0.1$	&	$0.028	\pm	0.007$	&	$0.0078	\pm	0.0006$	&	$176	\pm	4$	&	$51.65	\pm	0.01$	&	$51.97	\pm	0.26$	&	$9600	\pm	2900$	\\
971214	&	3.41	&	$2.3	\pm	0.11$	&	$0.010	\pm	0.004$	&	$0.0175	\pm	0.0012$	&	$107	\pm	6$	&	$52.69	\pm	0.02$	&	$52.62	\pm	0.27$	&	$29500	\pm	9100$	\\
980703	&	0.97	&	$2.6	\pm	0.12$	&	$0.147	\pm	0.056$	&	$0.0025	\pm	0.0005$	&	$181	\pm	9$	&	$51.38	\pm	0.02$	&	$51.16	\pm	0.28$	&	$5200	\pm	1700$	\\
990123	&	1.60	&	$16.6	\pm	0.24$	&	$0.015	\pm	0.005$	&	$0.0120	\pm	0.0005$	&	$267	\pm	3$	&	$52.74	\pm	0.01$	&	$52.49	\pm	0.26$	&	$9500	\pm	2900$	\\
990506	&	1.30	&	$22.2	\pm	0.27$	&	$0.011	\pm	0.004$	&	$0.0320	\pm	0.0080$	&	$280	\pm	30$	&	$52.64	\pm	0.01$	&	$52.94	\pm	0.29$	&	$13800	\pm	4600$	\\
990510	&	1.62	&	$10.2	\pm	0.2$	&	$0.012	\pm	0.003$	&	$0.0352	\pm	0.0014$	&	$74	\pm	2$	&	$52.54	\pm	0.01$	&	$52.65	\pm	0.26$	&	$14600	\pm	4400$	\\
991216	&	1.02	&	$82.1	\pm	0.5$	&	$0.0050	\pm	0.0020$	&	$0.0152	\pm	0.0003$	&	$250	\pm	3$	&	$52.95	\pm	0.01$	&	$52.68	\pm	0.27$	&	$5400	\pm	1600$	\\
000131	&	4.5	&	$1.8	\pm	0.2$	&	$0.0009	\pm	0.0004$	&	$0.0121	\pm	0.0018$	&	$183	\pm	15$	&	$52.86	\pm	0.05$	&	$53.08	\pm	0.28$	&	$57000	\pm	18500$	\\
 \enddata

\tablenotetext{a}{Peak flux for the brightest 256 ms time interval in $photon ~ s^{-1} cm^{-2}$ for 50-300 keV \citep{pac99}.}
\tablenotetext{b}{Lag in seconds as calculated from the peak of the cross correlation between BATSE channels 3 and 1 using data brighter than 10\% of the peak flux \citep{nmb00,nor02}.  These lags have been corrected by $1+z$ to the frame of the GRB.  The quoted one-sigma measurement uncertainties in the log of the lag should be added in quadrature with 0.35, the population dispersion, to get the effective uncertainty for luminosity determination.}
\tablenotetext{c}{Variability following the definition of \citet{frr00}; where $V$ is the normalized variance of the light curve around a $0.15 T_{90}$ box-smoothed light curve.  The quoted one-sigma measurement uncertainties in the log of the variability should be added in quadrature with 0.20, the population dispersion, to get the effective uncertainty for luminosity determination.}
\tablenotetext{d}{Photon energy (in keV) of the peak in the $\nu F_{\nu}$ spectrum as measured by \citet{mal95} with BATSE data.}
\tablenotetext{e}{The base-10 logarithm of the observed luminosity in
$erg ~s^{-1}$;  calculated as $4 \pi P \epsilon D_L^2$, where $\epsilon
\approx 1.7 \times 10^{-7} erg$ as the average energy of a photon in the
$50-300 keV$ range and $D_L$ is the luminosity distance which is taken
from $z$ for an assumed flat Universe with $\Omega_m = 0.3$ (with
equations 14, 15, 16, and 21 of Hogg 1999).  The calculated observed
luminosity depends on the adopted Hubble constant (here taken as
$65~km~s^{-1}Mpc^{-1}$); although the cosmology only depends on the {\it
shape} of the Hubble diagram and hence is independent of the Hubble
constant.}
\tablenotetext{f}{The base-10 logarithm for the combined luminosity in $erg ~s^{-1}$; calculated as the weighted average of the results of Eqs 1 and 2 (for which $\Omega_m = 0.3$ has been adopted).}
\tablenotetext{g}{The luminosity distance in Mpc calculated as $(L_{comb} / 4 \pi P \epsilon)^{0.5}$, for luminosity based on an $\Omega_m = 0.3$ calibration.}

\end{deluxetable}

\clearpage

\begin{figure}
\plottwo{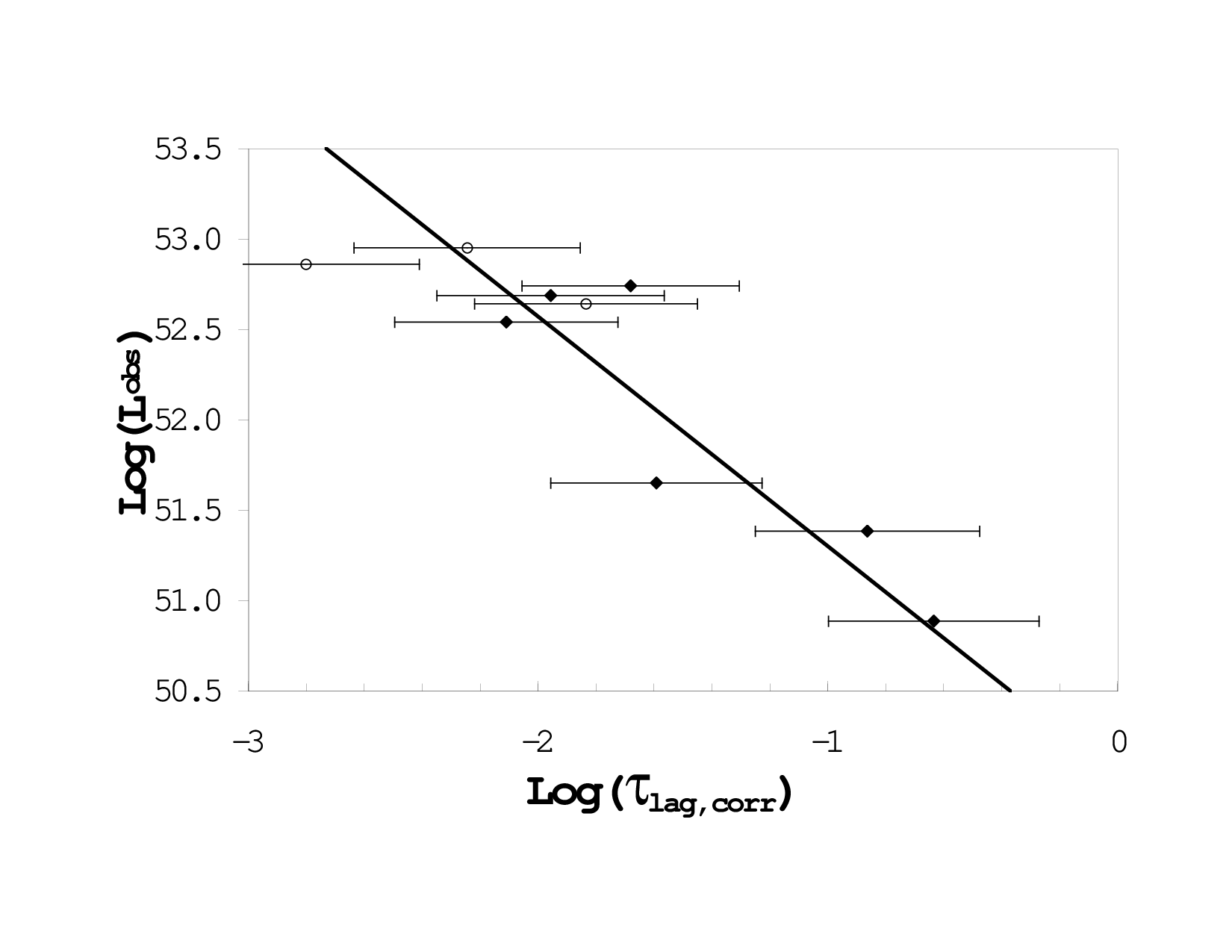}{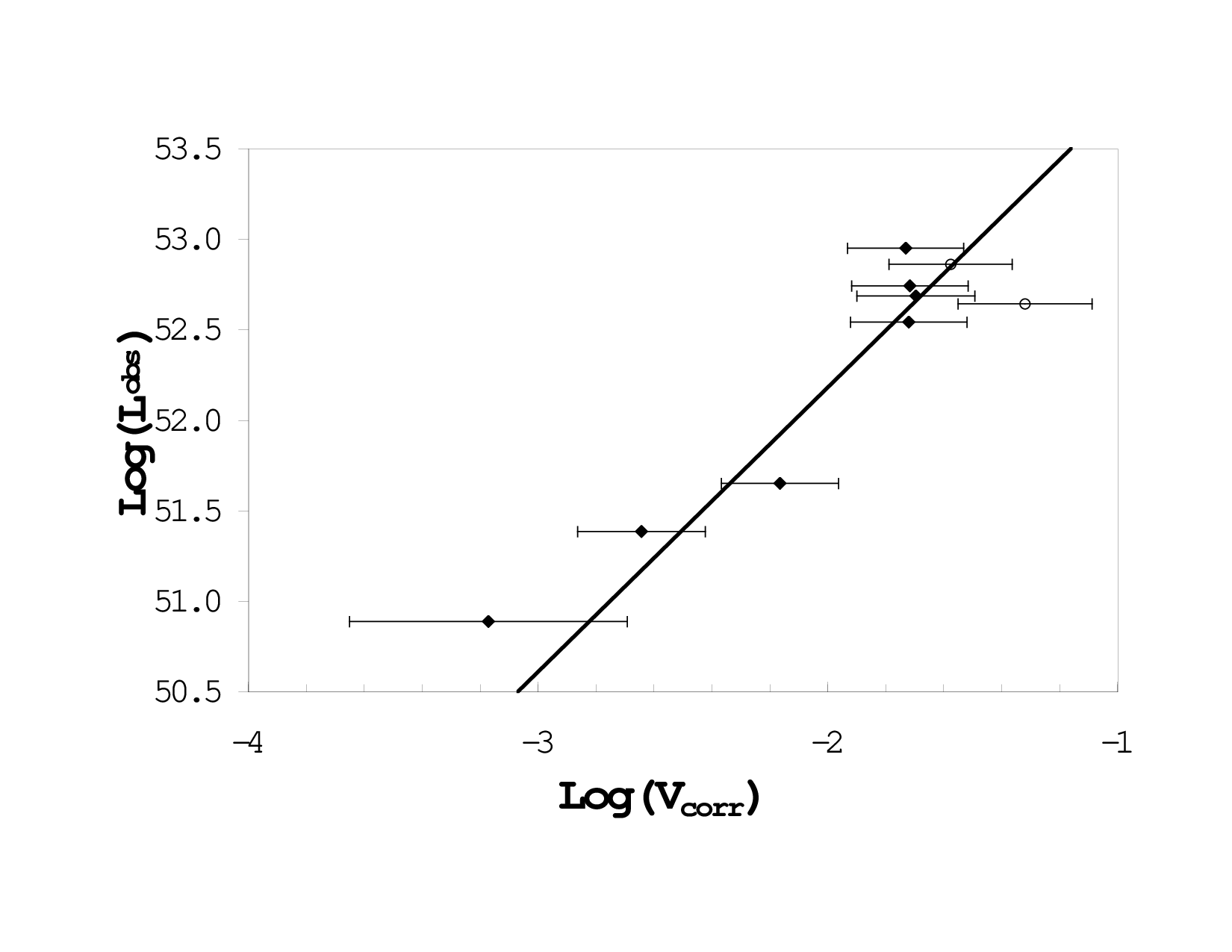}
\caption{Calibration curves for lag and variability.  Plots of the observed luminosity versus lag and variability can be used to calibrate the luminosity indicators.  Sufficient data are available for only 9 bursts with redshifts.  The plots here can be fitted to produce Eqs 1 and 2 (displayed as the straight lines in the two panels).  The displayed error bars include not only the observed measurement uncertainties but also the intrinsic scatter for the burst population of 0.35 and 0.20 (for lag and variability respectively) in logarithmic units, as determined by making the reduced chi-square of the fits close to unity. \label{fig1}}
\end{figure}

\clearpage 

\begin{figure}
\columnwidth=0.5\columnwidth
\plotone{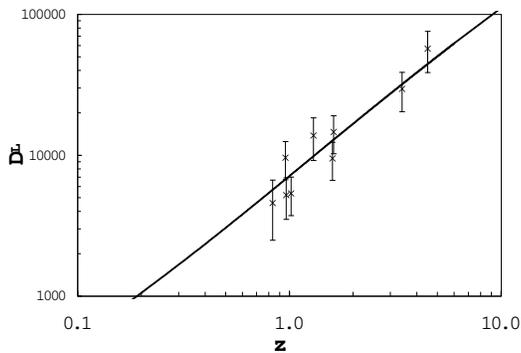}
\columnwidth=2\columnwidth
\caption{GRB Hubble Diagram.
This Hubble diagram is for 9 bursts from $0.84<z<4.5$.  With SWIFT, we should soon have $\sim 120$ bursts from roughly $0.1<z<10$, and this will allow a precision measure of the expansion history of the Universe to unprecedented high  redshifts.  The particular diagram above was made for luminosity indicators calibrated assuming a flat $\Omega_m = 0.3$ Universe (as in Figure 1 and Eqs 1 and 2) and the chi-square for the data versus the cosmological model (the curve in the figure) is 6.2.  The best fit cosmology can be found by minimizing the resulting chi-square as the cosmological model varies both the calibration curves and the resulting Hubble diagram.\label{fig2}}
\end{figure}

\end{document}